# The Barocaloric Effect:
# A Spin-off of the Discovery of High-Temperature Superconductivity

Albert Furrer

Laboratory for Neutron Scattering, Paul Scherrer Institut, CH-5232 Villigen PSI, Switzerland, and

SwissNeutronics AG, Bruehlstrasse 28, CH-5313 Klingnau, Switzerland

**Abstract:**
Some key results obtained in joint research projects with Alex Müller are summarized, concentrating on the invention of the barocaloric effect and its application for cooling as well as on important findings in the field of high-temperature superconductivity resulting from neutron scattering experiments.

## 1. Introduction

The magnetocaloric effect, *i.e.*, cooling by adiabatic demagnetization is a well-known technique in condensed matter physics [1]. The effect is a consequence of the variation in the total entropy of a solid by the magnetic field. In a first stage, an external magnetic field is applied isothermally (the system is in contact with a heat sink), thus reducing the magnetic entropy of the system. In a second stage, the magnetic field is removed adiabatically (the system is decoupled from the heat sink). In order to keep the entropy unchanged, the system is forced to reduce the temperature. Magnetic refrigeration has proven to be one of the most efficient cooling techniques in a wide range of temperatures up to room temperature and above [2].

Up to now, all magnetic refrigerators have suffered from the drawback of needing large magnetic fields of a few T in order to achieve cooling effects in the K range. In order to overcome this problem, Alex Müller proposed already in 1984 an alternative, namely to implememt adiabatic cooling by the application of pressure. More specifically, he proposed to make use of the pressure-induced structural phase transitions in $PrAlO_3$, which are accompanied by changes of the point symmetry of the $Pr^{3+}$ ions. This results in different splittings of the (2J+1)-fold degenerate ground-state multiplet by the crystal field, which governs the thermodynamic properties of the system at low temperatures. Hence the external pressure may well serve to change the magnetic entropy in the very same way as an external magnetic field does in the magnetocaloric effect (MCE). In analogy to the latter, Alex Müller gave the associated effect the name barocaloric effect (BCE). His idea was patented in several countries [3].

In what follows, I will discuss the developments and the results obtained by barocaloric cooling in co-operation with Alex Müller. The joint research efforts went considerably beyond the originally proposed mechanism based on structural phase transitions and included pressure-induced changes of a variety of phenomena such as magnetic phase transitions, the 4f-conduction-electron hybridization in Kondo systems, valence transitions, and spin fluctuations [4]. Finally, I will briefly summarize some of the key results obtained for high-temperature superconductivity by neutron scattering experiments jointly performed with Alex Müller.

## 2. The Barocaloric Effect (BCE)

### 2.1. Basic Principles

Structural phase transition in rare-earth (R) based perovskites usually occur as a function of temperature, but they can also be induced by the application of magnetic fields or pressure. If in one phase the R ions have a non-degenerate ground state and in the other a degenerate one, then cooling can be achieved under adiabatic application (or removal) of pressure (hydrostatic or uniaxial) due to the change of the entropy. The total entropy $S=S_m+S_e+S_p$ has magnetic, electronic, and phononic contributions, respectively. For rare-earth compounds and low temperatures, the magnetic entropy is the dominating contribution, which is determined by the (2J+1) split energy levels of the ground-state J-multiplet of the R ions:

$$S_m=-Nk_B\Sigma_i p_i \ln(p_i) , \quad p_i=\exp(-E_i/k_BT)/Z , \qquad (1)$$

where N is the total number of R ions in the system, $k_B$ the Boltzmann constant, $p_i$ the Boltzmann population factor of the energy levels $E_i$, and Z the partition function. The energy splittings due to the crystal-field interaction can be determined directly by the inelastic neutron scattering (INS) technique as exemplified in Fig. 1 for $Pr_{1-x}La_xNiO_3$ with different La concentrations x [5]. For x=0.15 the compound is in the orthorhombic Pbnm phase where the crystal-field interaction completely lifts the nine-fold degeneracy of the ground-state multiplet of the $Pr^{3+}$ ions, thus resulting in a singlet ground-state. For the rhombohedral R-3c phase corresponding to x=0.70, some of the crystal-field levels remain degenerate, and the ground state is a doublet. Fig. 2 shows the corresponding entropy curves calculated from Eq. (1) which exhibit a very different behavior over the whole temperature range and only merge together at very low temperatures [6].

Let us assume that the system is in the orthorhombic Pbnm phase at T=150 K, say at the point A marked in Fig. 2. We now perform the process of adiabatic pressurization. The system is isolated from its surroundings and an external pressure is applied such that the system transforms to the rhombohedral phase R-3c, thereby moving horizontally to the point B. The effect of this process is to lower the temperature significantly. For the next step the sample must be allowed to remain in contact with a heat sink, so that the process is isothermal. The sample is depressurized and moves vertically to the point C. In order to cool the system further down, the process of adiabatic pressurization and isothermal depressurization is repeated several times until the lowest possible temperature of about 200 mK is reached (see Fig. 2).

## 2.2. Structurally driven BCE

### 2.2.1. PrLaO$_3$

The perovskite compound PrAlO$_3$ exhibits a series of structural phase transitions from cubic to rhombohedral to orthorhombic to monoclinic and ultimately to tetragonal symmetry at temperatures of 1320, 205, 151, and 118.5 K, respectively [7-10]. The crystal-field level splittings of the Pr$^{3+}$ ions considerably change across these phase transitions [9]. For instance, in the rhombohedral phase (space group R-3c) the crystal-field ground state is a non-Kramers doublet, whereas in the low-temperature tetragonal phase (space group Pbnm) the crystal-field ground state is a singlet. The phase transitions at 205, 151, and 118.5 K can be suppressed by the application of uniaxial pressure along the (111)-direction, so that the rhombohedral phase can be stabilized at low temperatures. This opens the way for barocaloric cooling as described in the previous paragraph.

INS experiments were planned to investigate the pressure-induced effects on the magnetic excitation spectrum of PrAlO$_3$. For that purpose, Alex Müller and his technician W. Berlinger constructed a uniaxial pressure device for neutron scattering applications as described in Ref. [11]. The stress is generated in a hydraulic cylinder and transmitted through a plunger onto the single-crystal sample mounted in a cryostat. The assembly allows the application of a variable force up to $10^4$ N at temperatures between 2 and 300 K. Unfortunately, the experiments performed in the year 1985 turned out to be unsuccessful, since the single crystals always broke already at moderate pressures. Nevertheless, Alex Müller generously allowed to use the pressure device for other research projects. This opportunity was an extremely welcome extension of the sample environment for neutron scattering experiments at my laboratory and featured many successful applications such as the determination of (p,T) phase diagrams (see, *e.g.*, the example described in Ref. [11]).

### 2.2.2. $Pr_xLa_{1-x}NiO_3$

The discovery of high-temperature superconductivity temporarily pushed away Alex Müller's research interests from barocaloric cooling, but in the year 1991 – at a lunch during a conference in Kanazawa (Japan) – he showed me a preprint of a research work carried out by Jerry B. Torrance (IBM San Jose, USA) [12]. In his letter Jerry Torrance wrote: „Rumor spreads fast, especially when popular personalities are involved! I have heard that you are interested in the insulator-metal phase transitions we have found in rare-earth nickel compounds." Indeed, the $RNiO_3$ compounds turned out to be extremely suitable systems for barocaloric cooling. More specifically, the mixed compound $Pr_{1-x}La_xNiO_3$ undergoes a structural phase transition (SPT) from a high-temperature rhombohedral R-3c to a low-temperature orthorhombic Pbnm phase in the concentration range 0≤x≤0.7 [5,13]. The temperature $T_{SPT}$ is strongly dependent on the La concentration x: For x=0, we have $T_{SPT}$≈700 K, and $dT_{SPT}/dx$≈-1000 K. The SPT exhibits a peculiar temperature behavior as shown for $Pr_{0.66}La_{0.34}NiO_3$ in Fig. 3. The region in which the SPT occurs turns out to be very broad, *i.e.*, the two phases coexist in a rather large temperature range, thus $T_{SPT}$=361 K is defined as the temperature where the R-3c and Pbnm phases have equal weight. The application of hydrostatic pressure was found to shift $T_{SPT}$ to lower temperatures, typically by $dT_{SPT}/dp \approx -5$ K/kbar. This means that external pressure modifies the sample such that the volume fraction of rhombohedral symmetry is enhanced at the expense of the orthorhombic component. As mentioned above, the SPT is accompanied by a change of the degeneracy of the crystal-field ground state, so that barocaloric cooling becomes possible as illustrated in Fig. 2. Moreover, by varying the La concentration x, the working point for barocaloric cooling can be chosen at any temperature between 1 and 700 K.

The theoretical predictions of barocaloric cooling were verified in experiments performed for $Pr_{0.66}La_{0.34}NiO_3$ with $T_{SPT}$=361 K [14]. The setup for the pressure experiments is schematically shown in Fig. 4. The sample is enclosed in a cylindrical pressure cell of 5 mm diameter and 6 mm height, which allowed the *in situ* measurement of the sample temperature. The results are summarized in Fig. 5. At an initial temperature of 300 K and for hydrostatic pressures up to 15 kbar the compound is predominantly in the orthorhombic state (see Fig. 3). Through the application of pressure the compound gains elastic energy, *i.e.*, its entropy is lowered and the sample warms up. A similar situation is encountered at 400 K where the compound is predominantly in the rhombohedral state. At 350 K, however, the application of pressure results in a significant increase of the rhombohedral volume fraction, so that the compound experiences – in addition to the elastic heating – the magnetic cooling as described in the previous paragraphs, *i.e.*, we have a situation with two competing effects. As shown in Fig. 5(b), we observe a slight cooling of the compound for p=5 kbar, *i.e.*, the magnetic cooling exceeds the elastic heating. For p=10 kbar we observe an almost perfect counterbalance of the two competing effects. For p=15 kbar the elastic heating dominates the magnetic cooling.

The undesired effect of elastic heating observed in the above experiments can be drastically reduced by using single crystals and uniaxial pressure.

Subsequently we followed this route by studying the BCE in a series of suitable single crystals and extended the method to include – in addition to the structurally driven BCE demonstrated for $Pr_{0.66}La_{0.34}NiO_3$ – other sources of pressure-induced entropy changes. In particular, the magnetically driven BCE due to pressure-induced magnetic phase transitions implies an internal magnetic field resulting in large Zeeman splittings which in many cases are comparatively larger than by externally applied magnetic fields; the magnetically driven BCE may hence be considered the analogue to the MCE. Further, many rare-earth based Kondo systems show a distinct pressure dependence in the degree of 4f conduction electron hybridization and hence allow for a cooling. Finally, cooling can even be realized by the virtual exchange of the type of rare-earth ion via a pressure-induced valence transition, *e.g.*, $Eu^{2+} \rightarrow Eu^{3+}$. All these mechanisms have proven to lead to an effective cooling as summarized below.

*2.3. Magnetically driven BCE*

The rare-earth monopnictide CeSb orders antiferromagnetically below $T_N \approx 16$ K in various phases [15]. Uniaxial pressure along the [001] axis is known to increase $T_N$ by $dT_N/dp \approx 8$ K/GPa as shown in Fig. 6(a) [16]. In the paramagnetic state the six-fold degenerate ground-state multiplet of the $Ce^{3+}$ ion is split by the crystal field into a doublet and an excited quartet at 3.2 meV [17]. Below $T_N$ the crystal-field levels are further split by the molecular field due to the Zeeman effect giving rise to a strong decrease of the magnetic entropy as schematically shown in Fig. 6(b). There is a discontinuity of about $\Delta S \approx 2$ J/moleK when the first-order phase transition occurs. Barocaloric cooling can be implemented as follows: At point A the system is in the paramagnetic state at ambient pressure and an initial temperature $T_A > T_N(p=0)$. Next we apply pressure while in thermal contact with a heat sink. The system transforms isothermally into the ordered state, *i.e.*, vertically to point B, thereby releasing heat to the heat sink. Finally we isolate the system from the heat sink and release the pressure. The system transforms back into the paramagnetic state along an adiabatic line to point C lowering its temperature by ΔT. Fig. 7(a) shows the expected dependence of the cooling rate in function of the initial temperature $T_A$. The cooling is effective in the temperature range from $T_N(p=0)$ to $T_N(p>0)$, giving rise to a characteristic triangular shape of the cooling rate ΔT *versus* $T_A$. This theoretical prediction was nicely verified in experiments carried out for pressures up to 0.52 GPa. Fig. 7(b) exemplifies the results upon releasing an uniaxial pressure of 0.26 GPa [4,18,19]. Similar experiments were performed for HoAs [4,20].

## 2.4. BCE in a Kondo system

In rare-earth Kondo compounds it is generally found that the hybridization between the 4f and conduction electrons leads to a reduction of the magnetic entropy. In the metallic Kondo compound $Ce_3Pd_{20}Ge_6$ this reduction is significant, as the crystal-field ground state of the $Ce^{3+}$ ions is a quartet. The application of pressure is expected to influence the magnetic entropy twofold: (i) a reduction of the cell volume is generally found to increase the degree of hybridization; (ii) uniaxial pressure leads to a structural distortion, so that the quartet ground-state is split into two doublets. Both effects will reduce the magnetic entropy and allow barocaloric cooling as verified in Refs. [4,21].

## 2.5. Valence driven BCE

Eu compounds are especially well suited for a valence driven BCE, since the divalent magnetic $Eu^{2+}$ ion is larger than the trivalent nonmagnetic $Eu^{3+}$ ion, so that pressure can induce a transition $Eu^{2+}{\rightarrow}Eu^{3+}$ implying drastic changes in the magnetic entropy. More specifically, $Eu^{2+}$ ions with J=7/2 and S=0 have an eight-fold degenerate ground state, whereas $Eu^{3+}$ ions with J=0 have a singlet ground state, so that the magnetic entropy difference $\Delta S(Eu^{2+} \rightarrow Eu^{3+})=R \ln(8)$ is amongst the largest to be realized practically, giving rise to huge temperature changes $\Delta T \approx 10$ K for a single step in the BCE process.

Early in the new millennium, Alex Müller received a message from Frank Steglich that the compound $EuNi_2(Si_{1-x}Ge_x)_2$ under study at the Max-Planck Institute at Dresden might have the right potential for the BCE due to the large pressure dependence of the valence transition $T_V$ with $dT_V/dp=105$ K/GPa [22]. BCE experiments were performed for a polycrystalline sample with x=0.85 and $T_V$=38 (48) K for cooling (warming) and hydrostatic pressures up to 0.48 GPa [4,23]. The observed temperature changes $\Delta T \approx 1$ K per single step in the BCE process turned out to be an order of magnitude smaller than theoretically expected, which is due to the lack in adiabaticity of the experimental setup as well as to the use of a polycrystalline sample (suitable single crystals were not available).

## 3. Achievements in High-Temperature Superconductivity

### 3.1. Inhomogeneous Materials Character

Early in the nineties, Alex Müller put my attention to theoretical and experimental work going on at Stuttgart (University and Max-Planck-Institut für Festkörperforschung), which gave evidence for a phase separation in $La_2CuO_{4+x}$ compounds, *i.e.*, the co-existence of domains of essentially stoichiometric insulating $La_2CuO_4$ and domains of superconducting areas [24,25]. At the same time, we have been involved in neutron scattering studies of the crystal-field interaction in $RBa_2Cu_3O_x$ (R=rare earth) compounds in order to understand how the antiferromagnetic ordering of the $R^{3+}$ ions can co-exist with the superconducting state. These experiments revealed unusual features which later could be interpreted in terms of phase separation. Since the $R^{3+}$ ions are situated close to the $CuO_2$ planes where the superconducting carriers are located, the crystal-field interaction at the R site constitutes an ideal probe of the local symmetry and the charge distribution of the superconducting $CuO_2$ planes and thereby monitors directly changes of the carrier concentration induced by the oxygen nonstoichiometry x. More specifically, the lowest crystal-field excitation observed for $ErBa_2Cu_3O_x$ turned out to be decomposed into three individual transitions $A_i$ whose spectral weights distinctly depend on the oxygen content x [26]. Fig. 8 shows the fractional proportions $p_k$ of the transitions $A_1$, $A_2$, and $A_3$ which have maximum weight close to x=7.0, x=6.5, and x=6.0, respectively.

It is tempting to identify the transitions $A_1$, $A_2$, and $A_3$ by two local regions of metallic character ($T_c \approx 90$ K, $T_c \approx 60$ K) and a local region of semiconducting character, respectively, thereby providing evidence for the percolation mechanism of superconductivity in cuprates as proposed in Refs. [24,25] as well as for the two-plateau structure of $T_c$ in the compounds $RBa_2Cu_3O_x$. For x=6 the system is a perfect semiconductor. When x increases, oxygen ions are added into the chains, and holes are continuously transferred into the $CuO_2$ planes. By this mechanism the number of local regions with metallic character (associated with the crystal-field transition $A_2$) rises, which can partially combine to form larger regions. For a two-dimensional square structure the critical concentration for bond percolation is 50%. According to Fig. 8 the critical concentration is $x_2$=6.40, where the system undergoes a transition to the conducting state (with $T_c \approx 60$ K). Upon further increasing x, a second (different) type of metallic cluster (associated with the crystal-field transition $A_1$) is formed; these start to attach to each other, and at the percolation limit $x_1$=6.84 a transition into another conducting state (with $T_c \approx 90$ K) is induced.

## 3.2. Oxygen and Copper Isotope Effects

The concept which led Alex Müller to the discovery of superconductivity in the cuprates was the vibronic property of the Jahn-Teller effect. Therefore looking for an isotope effect for both oxygen and copper ions was an obvious task, and neutron crystal-field studies provide relevant information through the linewidth of the crystal-field transitions. The interaction with the charge carriers is by far the dominating relaxation mechanism. The corresponding intrinsic linewidth increases almost linearly with temperature according to the well-known Korringa law. In the superconducting state as well as in the pseudogap state, however, the pairing of the charge carriers creates an energy gap $\Delta$, thus crystal-field excitations with energy <$2\Delta$ do not have enough energy to span the gap, and consequently there is no interaction with the charge carriers.

We applied this method to the compound $HoBa_2Cu_4O_8$ for both oxygen isotope substitution with $T_c(^{16}O)$=79.0(1) K and $T_c(^{18}O)$=78.5(1) K [27] and copper isotope substitution with $T_c(^{63}Cu)$=79.0 K and $T_c(^{65}Cu)$=78.6(1) K [28]. Fig. 9 shows the temperature dependence of the intrinsic linewidth (HWHM) corresponding to the lowest-lying crystal-field transition at energy 0.6 meV. The linewidth is zero up to some tens K, then it increases almost linearly up to around 200 K, where a steplike enhancement into the normal state behavior according to the Korringa law occurs. The temperature T* associated with the steplike enhancement is identified as the temperature where the pseudogap opens. Our experiments gave evidence for large isotope shifts $\Delta T^*(O) \approx 50$ K and $\Delta T^*(Cu) \approx 25$ K. The corresponding isotope coefficients $\alpha^*$ defined by the relation $T^* \propto 1/M^{\alpha^*}$ (M is the mass of the O or Cu ion) turn out to be $\alpha^*(O)$=-2.2 and $\alpha^*(Cu)$=-4.9. Giorgio Benedek highlighted the latter coefficient in his lecture presented at an international symposium in Zurich in the year 2006 and suggested to call it the *Alfa Romeo number*, since Alex Müller is an enthusiastic driver of Alfa Romeo cars such as the model Alfa Romeo Montreal 4.9.

Subsequently this method was applied to the compound $La_{1.96-x}Sr_xHo_{0.04}CuO_4$ [29,30]. We found oxygen isotope coefficients $\alpha^*(O)$=-2.0, -1.3, -0.7, and -0.3 for x=0.11, 0.15, 0.20, and 0.25, respectively, but the copper isotope coefficient $\alpha^*(Cu)$ turned out to be zero within experimental error. This is in contrast to the case of $HoBa_2Cu_4O_8$, for which the large copper isotope shift was associated with a local copper mode, although the experiment did not provide direct information about the specific type of lattice mode involved. By comparing $\Delta T^*(Cu)$ for both $HoBa_2Cu_4O_8$ and $La_{1.96-x}Sr_xHo_{0.04}CuO_4$, it was possible to assign the copper mode to the umbrella-type mode [31], which is present in the former bilayer compound but not in the latter single-layer compound.

## 4. Concluding Remarks

Looking back to thirty years of collaboration I realize with pleasure that Alex Müller's inpact on my research work was significant. Due to his tremendous experience with perovskite compounds, a novel principle for cooling by the barocaloric effect became possible and could be demonstrated in experiments. I very much appreciated his invaluable contributions and sagacious advice in numerous discussions, often in my office under the inspiring influence of Peter's Flake (this is Alex Müller's favorite pipe tobacco to enjoy the fine art of smoking). I am also grateful to Alex Müller for actively promoting our achievements in neutron scattering studies of copper perovskites (inhomogeneous character of hole-doped cuprates, oxygen and copper isotope effects, nature of the gap function) in various viewpoints published in the literature (see, *e.g.*, Refs. [32,33]).

**Acknowledgments**

The author is indebted to the large number of colleagues for their participation in the experimental work as well as for numerous stimulating discussions.

**Figure Captions**

**Figure 1:**

Energy spectra of neutrons scattered from $Pr_{1-x}La_xNiO_3$ taken at T=10 K and modulus of the scattering vector Q=2.5 Å$^{-1}$ with x=0.15 (a) and x=0.70 (b) [5]. The top displays the crystal-field level sequence derived from the data, and the arrows mark the observed transitions. Some of the higher-lying crystal-field levels are not shown.

**Figure 2:**

Magnetic entropy calculated for the rhombohedral R-3c (dashed line) and orthorhombic Pbmn (full line) phases of $Pr_{1-x}La_xNiO_3$.

**Figure 3:**

Temperature dependence of the volume fraction of the rhombohedral R-3c phase (dashed line) and the orthorhombic Pbnm (full line) phases determined for $Pr_{0.66}La_{0.34}NiO_3$ by neutron diffraction [14].

**Figure 4:**

Schematic drawing of the pressure cell.

**Figure 5:**

Variation of the temperature of a $Pr_{0.66}La_{0.34}NiO_3$ sample upon applying a hydrostatic pressure up to 15 kbar for initial temperatures of 300 K (a), 350 K (b), and 400 K (c) [14].

**Figure 6:**

(a) p-T phase diagram of CeSb [16]. AF denotes the antiferromagnetic phases defined by the magnetic ordering wavevector **q**=(0,0,$q_0$). para marks the paramagnetic phase. (b) Schematic plot of the associated entropy curves at ambient (solid line) and at elevated pressure (dashed line).

**Figure 7:**

(a) Schematic plot of the expected cooling effect in CeSb deduced from the magnetic phase diagram of Fig. 6(a). (b) Measured cooling effect -ΔT in function of the initial temperature observed in CeSb upon releasing an uniaxial pressure of 0.26 GPa [4,18,19].

**Figure 8:**

Proportions $p_k$ of the lowest-lying crystal-field transitions $A_i$ of $ErBa_2Cu_3O_x$ as a function of the oxygen content x [26]. The full lines correspond to the probabilities $p_k$ to have 0≤k≤4 of the four oxygen chain sites nearest to the $Er^{3+}$ ion occupied. The dotted lines mark the critical limits for bond percolation.

**Figure 9:**

Temperature dependence of the intrinsic linewidth W (HWHM) corresponding to the lowest-lying crystal-field transition in $HoBa_2Cu_4O_8$ for different oxygen and copper isotopes [27,28]. The lines denote the linewidth in the normal state calculated from the Korringa law.

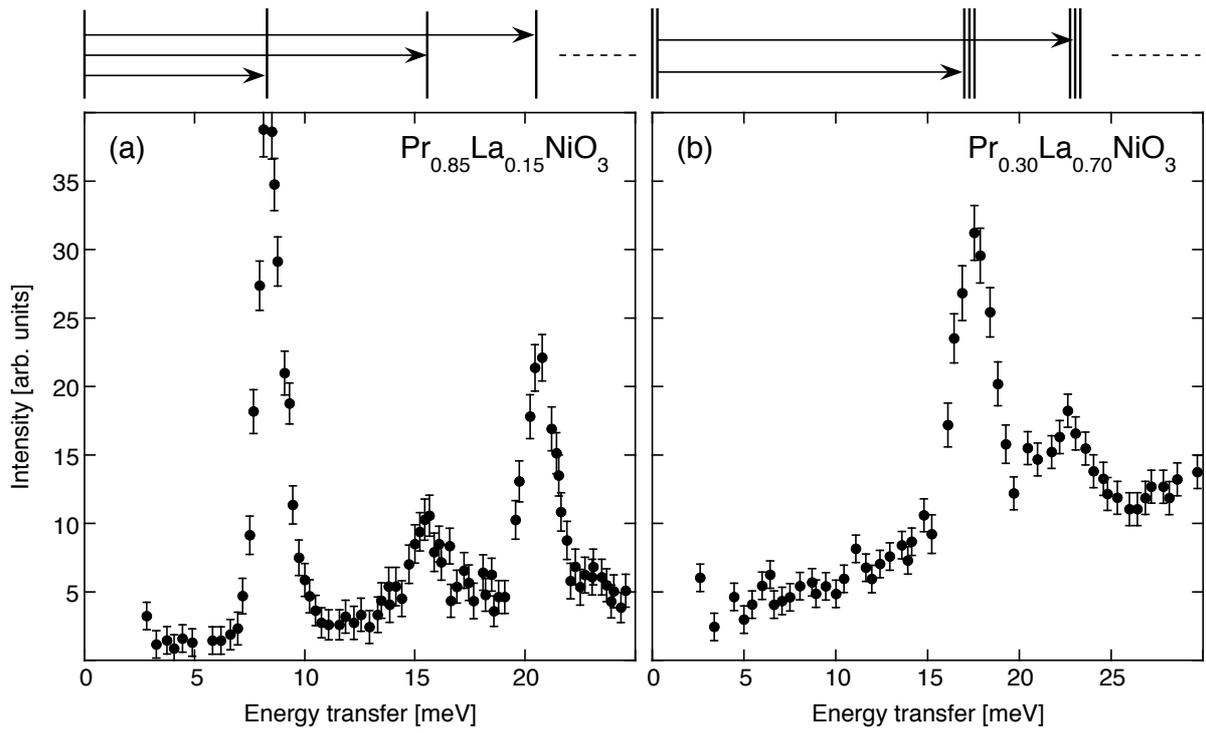

**Figure 1:**

Energy spectra of neutrons scattered from $Pr_{1-x}La_xNiO_3$ taken at T=10 K and modulus of the scattering vector Q=2.5 Å$^{-1}$ with x=0.15 (a) and x=0.70 (b) [5]. The top displays the crystal-field level sequence derived from the data, and the arrows mark the observed transitions. Some of the higher-lying crystal-field levels are not shown.

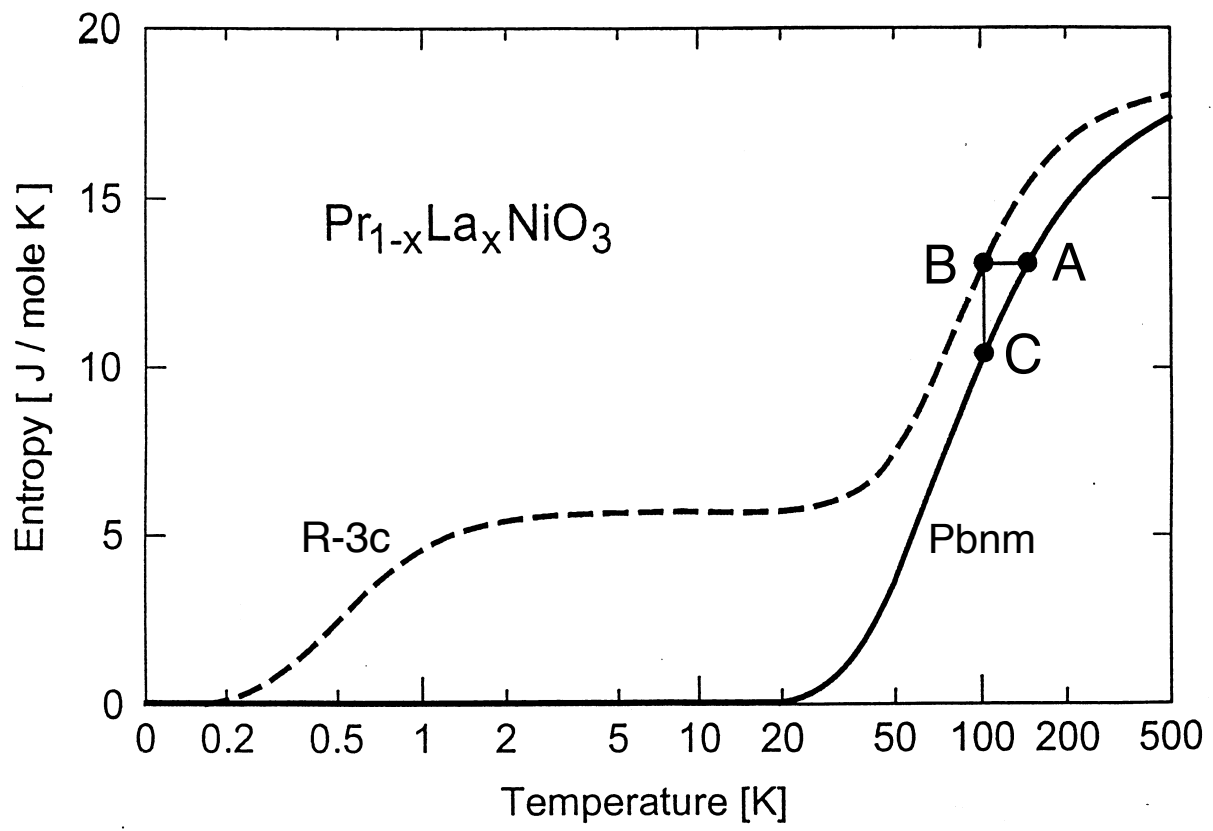

(dashed line) and

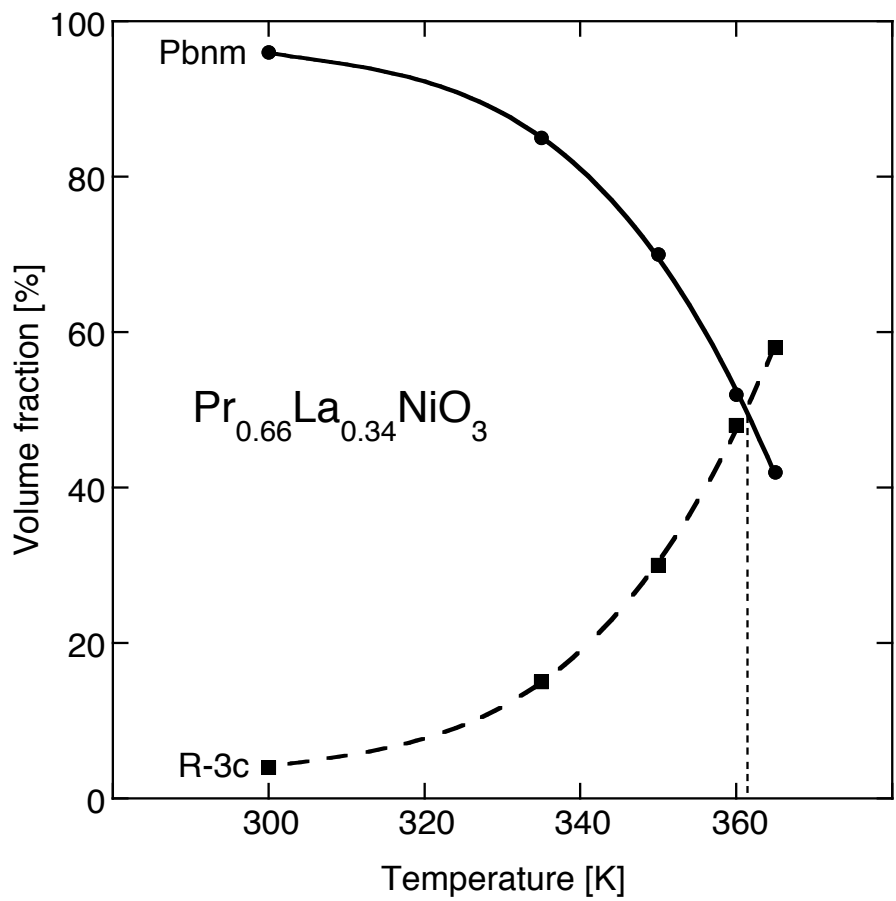

**Figure 3:**

Temperature dependence of the volume fraction of the rhombohedral R-3c (dashed line) and the orthorhombic Pbnm (full line) phases determined for $Pr_{0.66}La_{0.34}NiO_3$ by neutron diffraction [14].

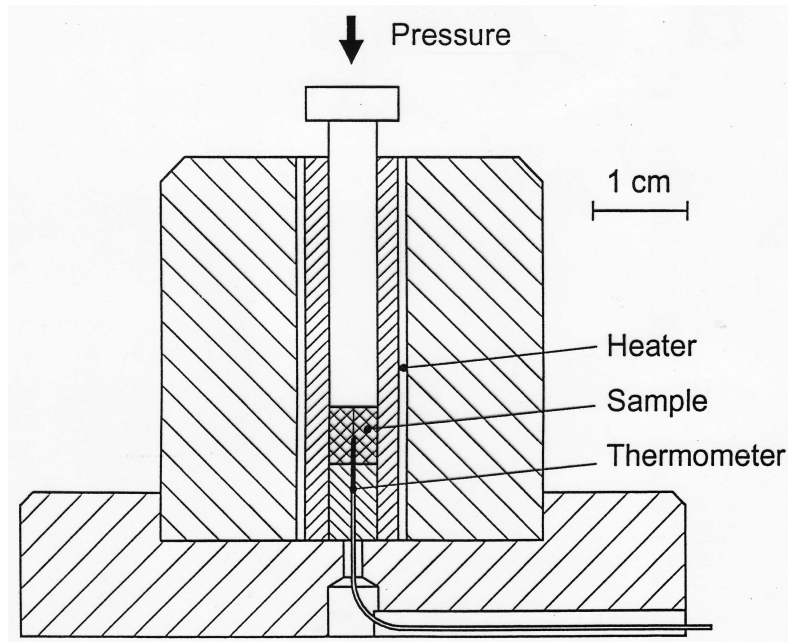

**Figure 4:**

Schematic drawing of the pressure cell.

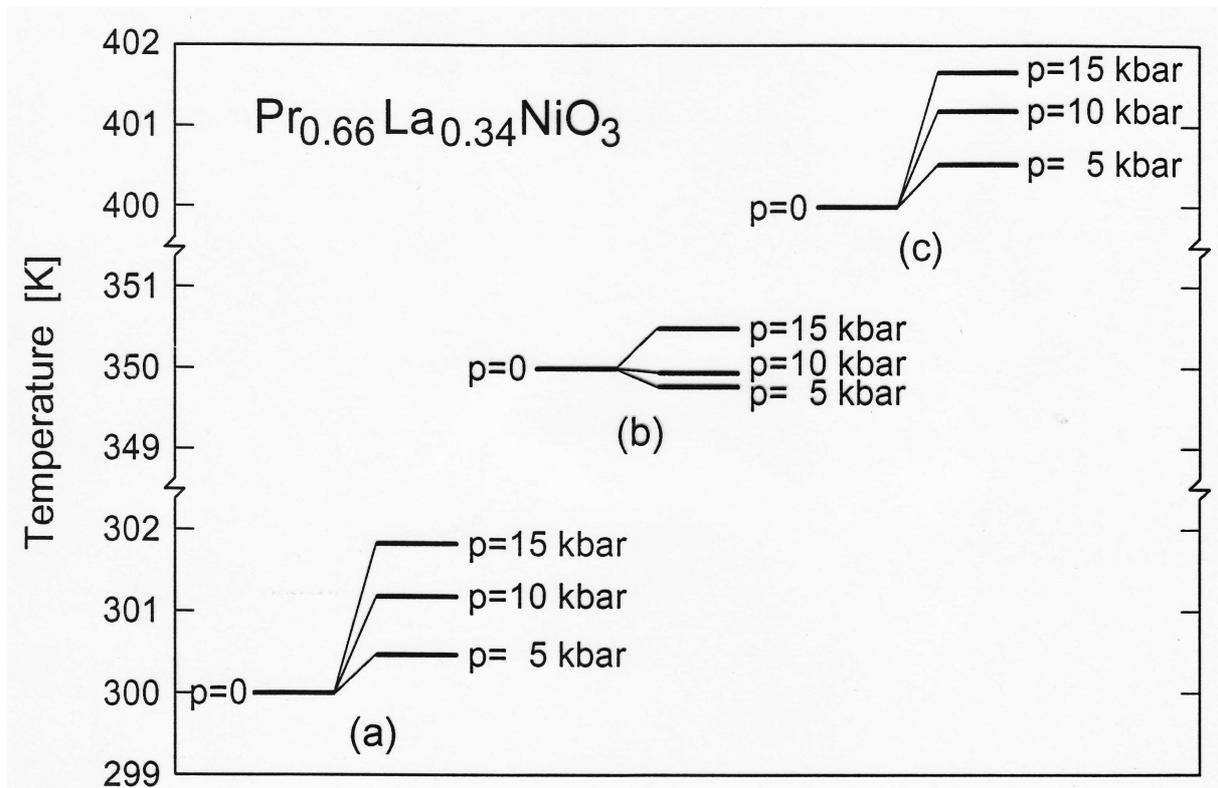

**Figure 5:**

Variation of the temperature of a $Pr_{0.66}La_{0.34}NiO_3$ sample upon applying a hydrostatic pressure up to 15 kbar for initial temperatures of 300 K (a), 350 K (b), and 400 K (c) [14].

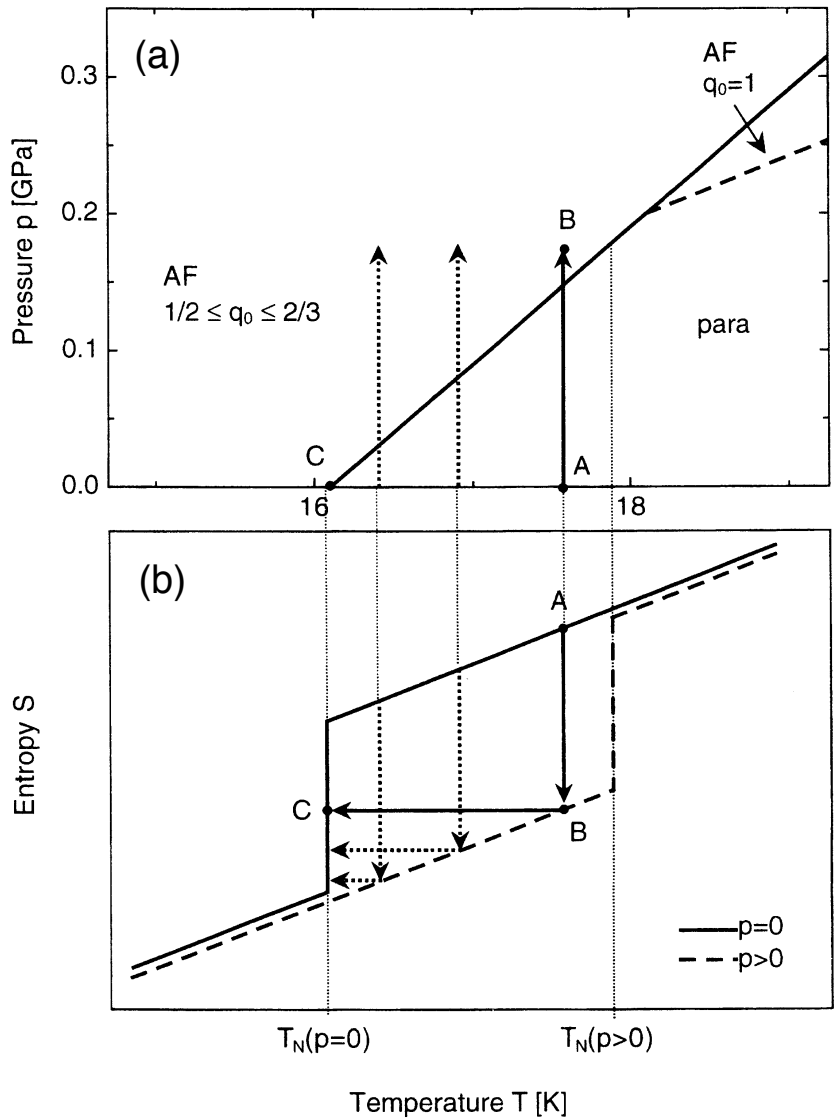

**Figure 6:**

(a) p-T phase diagram of CeSb [16]. AF denotes the antiferromagnetic phases defined by the magnetic ordering wavevector $\mathbf{q}=(0,0,q_0)$. para marks the paramagnetic phase. (b) Schematic plot of associated entropy curves at ambient (solid line) and at elevated pressure (dashed line).

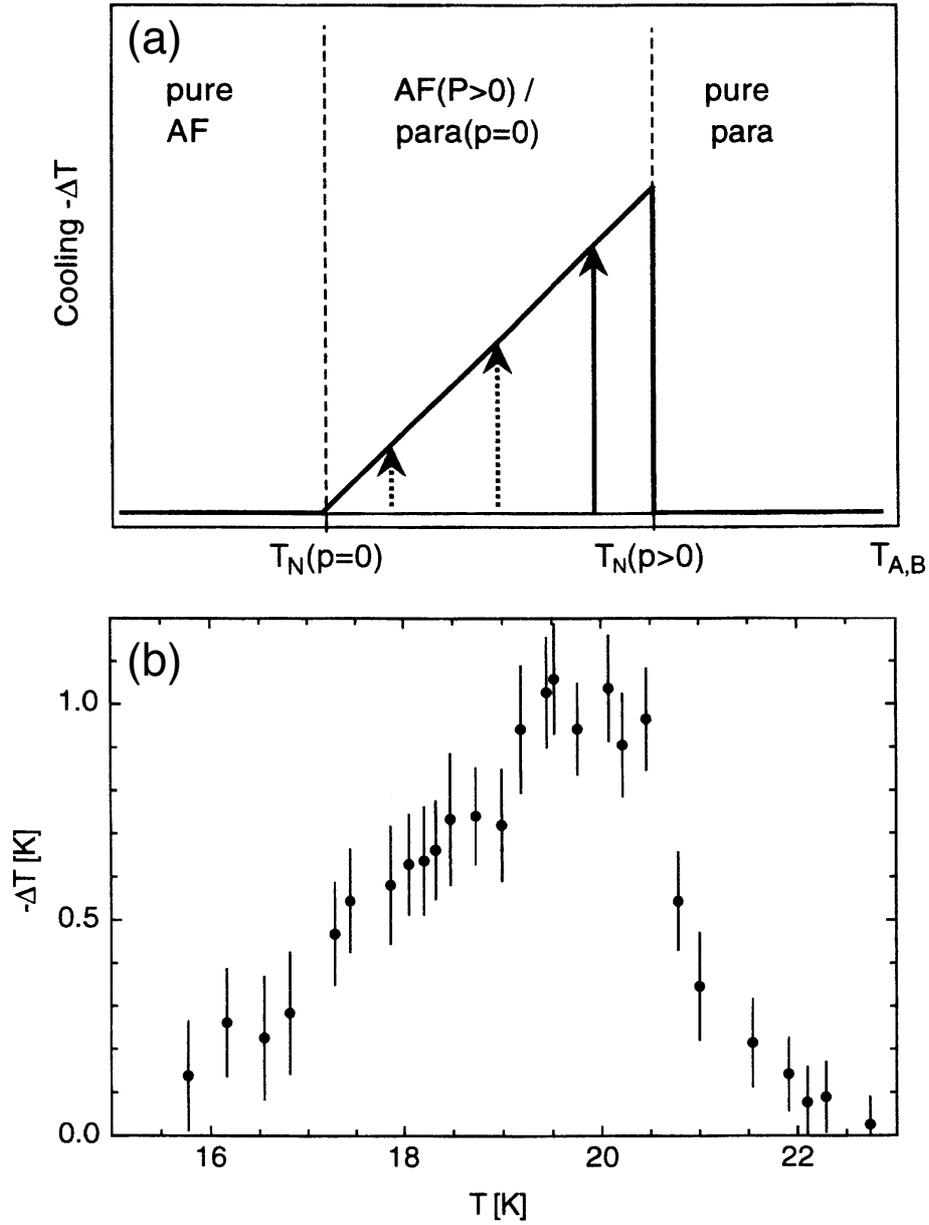

**Figure 7:**

(a) Schematic plot of the expected cooling effect in CeSb deduced from the magnetic phase diagram of Fig. 6(a). (b) Measured cooling effect -ΔT in function of the initial temperature observed in CeSb upon releasing an uniaxial pressure of 0.26 GPa [4,18,19].

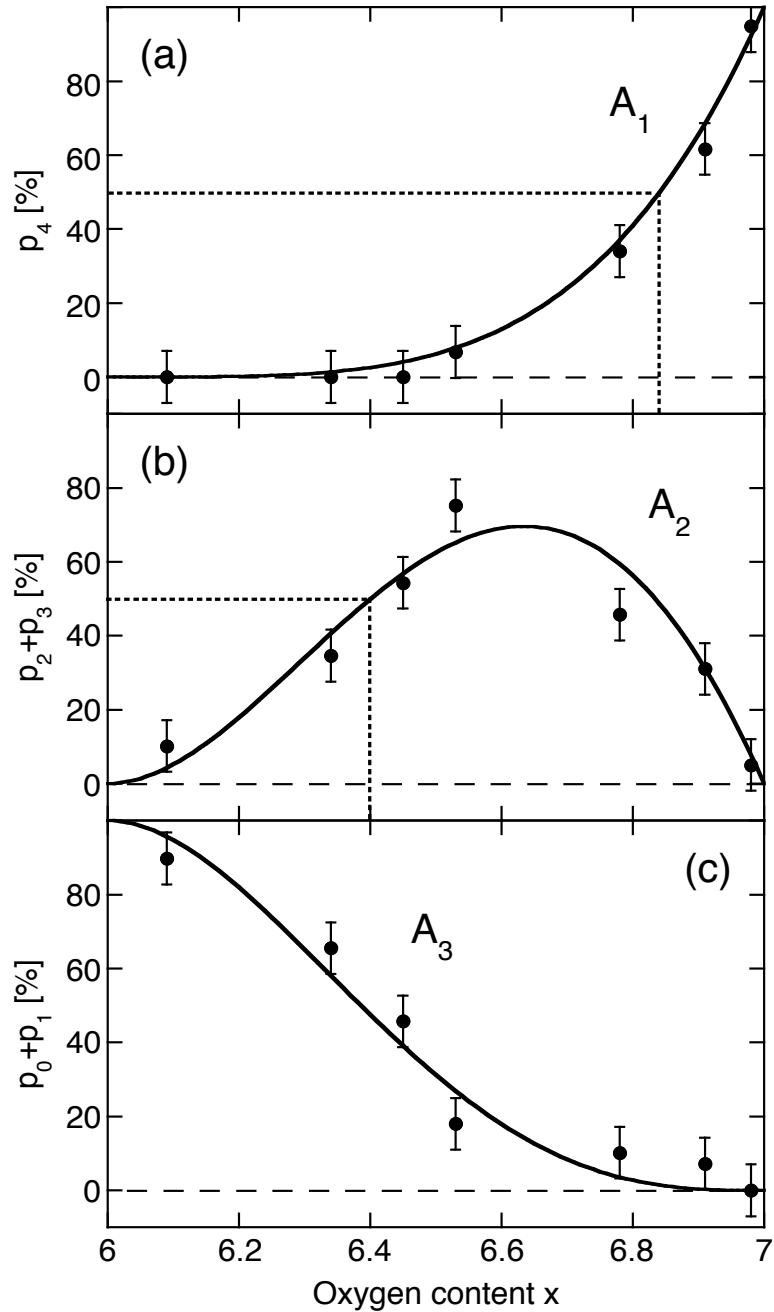

**Figure 8:**

Proportions $p_k$ of the lowest-lying crystal-field transitions $A_i$ of $ErBa_2Cu_3O_x$ as a function of the oxygen content x [26]. The full lines correspond to the probabilities $p_k$ to have $0 \leq k \leq 4$ of the four oxygen chain sites nearest to the $Er^{3+}$ ion occupied. The dotted lines mark the critical limits for bond percolation.

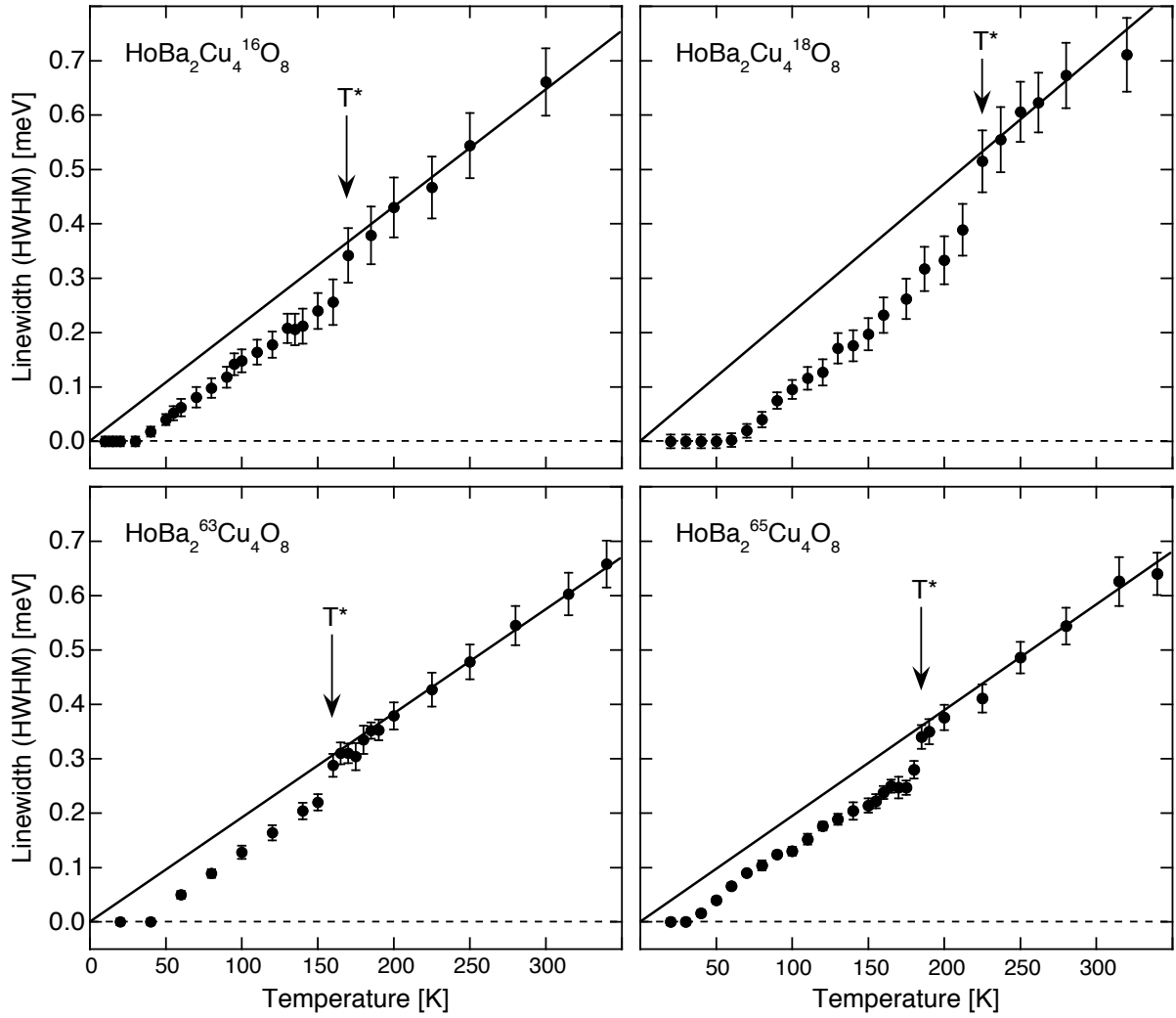

**Figure 9:**

Temperature dependence of the intrinsic linewidth W (HWHM) corresponding to the lowest-lying crystal-field transition in $HoBa_2Cu_4O_8$ for different oxygen and copper isotopes [27,28]. The lines denote the linewidth in the normal state calculated from the Korringa law.